# Simulating Special but Natural Quantum Circuits

Richard J. Lipton[†]    Kenneth W. Regan[‡]    Atri Rudra[*][‡]

[†]College of Computing,
Georgia Institute of Technology
`rjl@cc.gatech.edu`

[‡]Department of CSE,
University at Buffalo (SUNY),
`{regan,atri}@buffalo.edu`

**Abstract**

We identify a sub-class of BQP that captures certain structural commonalities among many quantum algorithms including Shor's algorithms. This class does not contain all of BQP (e.g. Grover's algorithm does not fall into this class). Our main result is that any algorithm in this class that measures at most $O(\log n)$ qubits can be simulated by classical randomized polynomial time algorithms. This does not dequantize Shor's algorithm (as the latter measures $n$ qubits) but our work also highlights a new potentially hard function for cryptographic applications.

Our main technical contribution is (to the best of our knowledge) a new exact characterization of certain sums of Fourier-type coefficients (with exponentially many summands).

[*]Supported by NSF CAREER grant CCF-0844796.

# 1 Introduction

One of the key problems in complexity theory is to determine the power of the complexity class BQP. Recall that this is the set of languages accepted by uniform polynomial size quantum circuits with bounded two-sided error. It is essentially the quantum version of the complexity class BPP. Just as BPP corresponds to what is feasible on a classical computer with randomness, BQP corresponds to what is feasible on a quantum computer. Whether BQP = BPP is a simple and focused way of asking, are quantum computers more powerful than classical ones? Whether actual quantum computers are constructed soon or in the distant future, this is a fundamental question.

Excitement has come from the discovery quantum algorithms that can out-perform classical algorithms by an exponential or at least quadratic amount. The most famous are Shor's factoring and discrete logarithm quantum algorithms [Sho97] and Grover's search algorithm [Gro97], plus real time simulation of many-body problems and other natural (quantum) processes as motivated by Feynman [Fey82]. Since these seminal results there have been many many new algorithms for problems in number theory, combinatorics, communication, and information theory that have applied quantum methods, as well as lower bounds for other problems.

The statement that quantum algorithms can out-perform classical algorithms needs to be examined carefully. In some cases this can be proved unconditionally, provided the algorithms are restricted. For example, it can be shown that provided both kinds of algorithms are restricted to *oracle access* to a boolean function, Grover's algorithm indeed is faster than any classical algorithm (see [DH09]). These are a pretty and important class of lower bounds, but without such restrictions we are still in the dark—we cannot unconditionally yet prove that quantum polynomial time is not just polynomial time, with-or-without bounded-error randomness. There are other approaches besides attempting direct lower bounds on BQP:

1. What are upper bounds for classical simulation of BQP?
2. What restricted cases allow feasible classical simulation?

Neither idea is new. Of course the former necessarily involves upper bounds for factoring. The latter has a long pedigree as described below, but first we single out two papers that are closest to ours in technical domain. DiVincenzo and Terhal [DT04] and Fenner et al. [FGHZ05, FFG$^+$06] studied constant-depth quantum circuits. These give polynomial-time classical simulations when either quantum fan-out is restricted ([DT04]) or success is highly amplified ([FGHZ05]), also involving restrictions on availability or use of *ancilla* qubits. Our paper studies special kinds of constant-depth circuits (some technically log-depth in their modeling) that have structures used by the above algorithms, but liberalizes the success probability, and focuses instead on the number of qubits measured to achieve it.

For upper bounds, Valiant appears to be the first to note that this class is easily contained in PSPACE. A much deeper result is that of Aldeman and DeMarrais and Huang [ADH97], who showed that BQP ⊆ PP. Since then there have been intense attempts to understand where quantum polynomial time resides. There are plausible conjectures that it is not contained in the polynomial time hierarchy, PH, but this is unproved (see [Aar10, FU10, AA11]). Since the classical analog, BPP, is contained in the hierarchy thanks to the beautiful results of [Lau83, Sip83], this would give evidence that quantum is different. Of course it would not be an unconditional proof since it still is possible—even if "unlikely"—that polynomial time could equal PSPACE, which would collapse many of our interesting complexity classes.

Serious work on simulating quantum computations by classical polynomial-time algorithms is often regarded as beginning with the Gottesman-Knill theorem [Got98] (see also [AG04, AB06, Joz08, vdN09, vdN10]), which shows that uniform BQP circuits restricted to a certain set of gates accept only languages in BPP.

Valiant, later, attempted to understand the power of quantum algorithms in a famous paper that introduced the concept of *matchgates* [Val06]. Roughly he viewed quantum algorithms as quantum circuits, and allowed only special two-bit gates that were characterized by $4 \times 4$ matrices in which the sixteen entries obeyed a set of algebraic relationships. For these circuits he was able to show that they did have classic polynomial time simulations. This was done by a brilliant insight that showed that these circuits were related to certain special types of determinants.

Subsequently, polynomial-time simulations have revolved around Valiant's theory of matchgates [Val06, CL07, Val08, Val10, GLV11] (cf. [CL08, CL09, CLX10]). They prove that any quantum circuit based on matchgates uses at most logarithmic qubits, and thus can be described by polynomial sized matrices. This explains in another way why they are efficiently simulated by classic methods: the unitary matrices are not too large to write down. Recall in a general quantum algorithm the matrices are usually exponential in size and no direct simulation is possible. Indeed even Valiant's initial observation that BQP is contained in PSPACE requires that the exponentially large matrices be handled in an "implicit manner."

**Quantum Background.** We now briefly summarize the concepts from quantum computing that are required to understand this paper. We talk about everything in terms of matrix-vector operations as in [For03], but allow more-general measurements in the standard basis.

The state of a quantum algorithm on $n$ qubits will be represented by a vector $x \in \mathbb{R}^N$, where $N \stackrel{def}{=} 2^n$. A quantum algorithm can be described as $x$ multiplied by a sequence of $N \times N$ *unitary* matrices $U_1, \ldots, U_m$ followed by a measurement (which can be followed by some classical computation). A unitary matrix preserves the $\ell_2$ norm of a vector; i.e., for any $y \in \mathbb{R}^N$ and an $N \times N$ unitary matrix $U$, we have $\|Uy\|_2 = \|y\|_2$. If the unitary matrix is a permutation then it corresponds to a deterministic (classical) computation. Finally, given $\hat{x} = U_m \cdot U_{m-1} \ldots U_1 \cdot x$, a quantum algorithm does some measurements. In particular, given any subset $R \subseteq [n]$ and an assignment $b \in \{0,1\}^{|R|}$, a measurement of the qubits $R$ to check whether the bits are equal to the assignment $b$ implies the computation of the quantity $\sum_{i \in S} |\hat{x}_i|^2$, where $S \subseteq [N]$ is defined as follows. For every $i \in [N]$ assign to it a unique vector $v(i) \in \{0,1\}^n$. Then $S$ is the set of all $i$ such that $v(i)$ projected down to the indices in $R$ is exactly the vector $\bar{b}$, (where $\bar{b}$ is $b$ with each bit flipped).

## 1.1 Our Approach

Our approach is based on an observation about many of the known quantum algorithms. These include the Deutsch-Jozsa algorithm [Deu85, DJ92, CEMM98], Simon's algorithm [Sim97], and Shor's algorithm [Sho97], *all share a structural property*. We will define the property in a moment, but for now let's call them $D^3$-algorithms, $D$ for Deutsch and $D^3$ for describable ○ deterministic ○ decomposable.

The property of being a $D^3$-algorithm is based solely on the structure of the quantum circuit. It is used by many other algorithms, even beyond the ones cited above. However, not all quantum algorithms have this structure: the most important one that does not is the quantum search algorithm of Grover. We believe this is an important point: if the class of algorithms we plan to study included all quantum algorithms we would most likely not make progress. But by limiting our research to a subclass we believe that progress is possible. Further, the fact that so many important algorithms have this structural property also means that the class is important.

Our main result is the following:

**Theorem 1.1.** *Any polynomial time $D^3$-algorithm that measures at most $O(\log n)$ bits can be simulated by a randomized polynomial time algorithm.*

Since this family of simulated algorithms includes Shor's factoring algorithm, does this mean that we can factor? No. The issue is the interplay between measurement and success probability. We note already forms of this issue that limits the polynomial-time simulation of Fenner et al. [FGHZ05]. This paper simulates in classical deterministic polynomial time any depth-$d$ quantum circuit $C$ with the property that for some given (but arbitrary) set $R$ of output qubits, either

- $R$ is measured to be in the all-zero state with probability $\geqslant 1 - \delta$, or
- $R$ is measured to be in that state with probability $\leqslant \varepsilon$,

provided $\delta \leqslant (1-\varepsilon)/2^{2d}$. This is a wide separation—in particular, it requires the probability in the "accept" case to be close to 1. There has apparently not been much progress in closing it [FG11]. For contrast, our result holds under the standard $(1/3, 2/3)$ separation, and versions allow for moderate changes either to the separation or the number of qubits by which it is achieved. Note that $R$ corresponds to a much larger set $S$ of co-ordinates in the underlying $N$-dimensional Hilbert space, where $N = 2^m$ in the case of $m = n^{O(1)}$ qubits. The number $m - n$ of *ancilla* qubits allowed to the circuits on inputs of size $n$ is also an issue, as shown in the followup paper [FFG+06] and related to the use of *fan-out* gates in [DT04].

**Technical Contributions.** We make, we believe, two main technical contributions. The first is the exact characterization of certain sums. In particular, given a unary matrix $M$ with certain properties and any vector $\mathbf{x}$, we exactly characterize the $\ell_2$ norm of certain sub-vector of $M\mathbf{x}$ (which correspond to certain quantum measurements) in terms of components of $\mathbf{x}$. To the best of our knowledge, this result is new.

For some instances of $D^3$ algorithms, the matrix $M$ turns out to be a Fourier transform (over $\{-1, 1\}$ or the reals). It is natural to wonder if existing sub-linear time algorithms ([GGI+02, AGS03, Aka10, Iwe08]) to compute Fourier coefficients could be useful in our context. In Section 5.1, we argue that these algorithms (at least as a black box) do not lead to an efficient simulations of the $D^3$ algorithms, while our exact characterization works.

We believe that this exact characterization is surprising, and potentially of use elsewhere in quantum theory as well as elsewhere in theory in general. In hindsight perhaps the exact characterization is simple, but initially we had an approximate lemma. Only later did we realize that the value we needed could be written down in closed form.

The other technical contribution is the notion of succinctness that we introduce. This notion based on the census idea seems new. While it is created to work perfectly with permutation matrices, it will likely be useful elsewhere in complexity theory. There are many studies of notions of succinctness and a new one could certainly lead to new results.

**A Cryptographic Implication.** The quantum part of Shor's algorithm actually does not compute the factorization. It computes an intermediate function (by measuring $n$ qubits) and then computes the factors from that output using ideas from continued fractions. Our algorithm only measures $O(\log n)$ bits. If we cannot push our positive results to measure $n$ qubits, then this points to potentially new hard problem that could be useful for cryptographic applications. In particular, the (potentially hard) problem would be: given say the measurement of $r(n)$ qubits from Shor's algorithm, compute the factorization. In fact by our result, when $r(n) = O(\log n)$, this problem is at least as hard as factoring (which of course is a widely used one-way function in practice). We plan to expand on this question in the final version of the paper.

**Organization.** We start with some preliminaries in Section 2. We then formally define our subclass of BQP in Section 3. We prove our main technical result in Section 4 and then use it to complete our proof of Theorem 1.1 in Section 5. We conclude with some open questions in Section 6.

## 2 Preliminaries

All the definitions we use are standard. We usually use lower case letters for vectors and upper case for matrices. If $x$ is a vector, then $x_k$ is the $k$-coordinate of the vector; if $A$ is a matrix, then $A_{ij}$ is the entry in position $(i, j)$. All our vectors and matrices are in Euclidean space $\mathbb{R}^\ell$ for some $\ell$. The inner product of two vectors is $(x, y)$. The 2-norm is $||x||_2$. If $x$ is a vector then $x^T$ is its transpose. Also an elementary vector is just a vector with one 1 and all the rest of its coordinates 0. If $x$ and $y$ are $n$-bit vectors, then $x \cdot y$ is their boolean inner product

$$x_1 y_1 \oplus \cdots \oplus x_n y_n.$$

The notions from quantum computing and complexity theory also are standard. As usual BQP is the complexity class of bounded error quantum polynomial time. The complexity class PH is the polynomial time hierarchy.

Some special notation. We always use $N$ to denote $2^n$. Usually $n$ is the number of qubits, which often involves padding the input size "$n_0$" with additional qubits, but when $n$ is linear or even polynomial in $n_0$ we can often ignore the distinction. For each quantum circuit $Q$ there is a corresponding unitary matrix $U_Q$ that corresponds to the action of the circuit.

All vectors are in $\mathbb{R}^N$, although all results generalize to complex vectors. We use $||v||_2$ for the $\ell_2$ norm of a vector and $||v||_\infty$ for the maximum norm.

We consider vectors and matrices whose entries come from a fixed finite set $F$. For a vector $v \in F^N$ and index $i \leqslant N$ the *census* $C_v(i)$ is the set of pairs $(a, |\{j \leqslant i : v_j = a\}|)$ giving the number of previous elements that equal $a$. If $v$ is a binary string then we can set $C_v(i) =$ the number of 1's in $v(1 \ldots i)$. For a matrix $A$, the census $C_A(i, j)$ similarly counts entries in row-major order.

**Definition 2.1.** *A vector or matrix of dimension $N = 2^n$ is* **succinct** *if its census is computable by a circuit of size bounded by a fixed polynomial in $n$. A subset $S$ of $[N]$ is succinct if its indicator string is succinct.*

Certainly these definitions imply the ability to compute the entries $v(i)$ or $A(i, j)$ themselves. The reason for the stronger census definition is to allow binary search:

**Lemma 2.2.** *(a) If $D$ is a succinct $N \times N$ permutation matrix, then there is a fixed polynomial size circuit that given $i$ computes $d(i)$ where $d : [N] \to [N]$ is the permutation computed by $D$.*

*(b) If $S$ and $S'$ are succinct sets of the same cardinality, then there is a succinct permutation $d$ that carries $S$ onto $S'$.*

*(c) The product of a succinct matrix and a succinct permutation matrix is succinct.*

*Proof.* For (a), do binary search on row $i$ implicitly using the row-major string. For (b), given any $i$ we can first test $i \in S$. If so, then use binary search to compute the census $C_1(i)$, namely the rank of $i$ in $S$, and use binary search again to find the element of the same rank in $S'$. If not, apply the same reasoning to the complements of $S$ and $S'$. Finally for $(c)$, to compute entry $(i, j)$ of $A' = DA$, it suffices to compute $A'(i, j) = A(d(i), j)$, since the previous $i - 1$ rows contribute $i - 1$ to the row-major order census. □

**Lemma 2.3.** *If $D$ is the matrix of a quantum circuit of fixed polynomial size with only deterministic gates drawn from a finite set of bounded-arity gates, then $D$ is succinct.*

*Proof.* Every elementary quantum gate defines a succinct permutation matrix, and succinctness of products follows from Lemma 2.2(c). To be careful with concrete versus asymptotic notation, if we have a $p(n)$-fold product $D_1 \cdots D_{p(n)}$ of permutation matrices each of whose succinctness is witnessed by circuits of size $q(n)$, then the composition of the permutations is computed by circuits of size $O(p(n)q(n))$, so the witness for its succinctness stays bounded by a fixed polynomial. □

We will say that we can compute a quantity $q$ to within an *additive error* $\varepsilon$ if we can compute $q'$ so that
$$|q - q'| \leqslant \varepsilon.$$

We will be interested in the $\ell_2^2$ norm of sub-vectors:

**Definition 2.4.** *Let $\mathbf{x} = (x_1, \ldots, x_N) \in \mathbb{C}^N$ and let $S \subseteq [N]$. Then let*
$$\mu_S(\mathbf{x}) = \sum_{i \in S} |x_i|^2.$$

Finally, we will need the definition of the count of a given pattern in "folded" vectors:

**Definition 2.5.** *Let $r \geqslant 1$ be an integer that divides $N$ and let $\mathbf{v} = (v_1, \ldots, v_r) \in \mathbb{C}^r$. Then for any $\mathbf{x} = (x_1, \ldots, x_N) \in \mathbb{C}^N$:*
$$\#_{\mathbf{v}}(\mathbf{x}) = |\{i \in [N/r] | v_j = x_{j \cdot N/r + i} \text{ for every } j \in [r]\}|.$$

## 3 The Structural Property

In this section we will explain the property that defines the class of $D^3$ quantum algorithms. Consider a polynomial size quantum circuit that consists of three stages: the $\mathcal{A}$, the $\mathcal{D}$, and the $\mathcal{B}$—see Figure 1. This is a $D^3$-circuit provided:

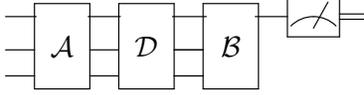

Figure 1: Our quantum circuit model

1. The circuit $\mathcal{A}$ defines a **succinct** unitary matrix, namely one with a special small circuit description.

2. The circuit $\mathcal{D}$ consists of only deterministic gates.

3. The circuit $\mathcal{B}$ defines a **decomposable** type unitary matrix.

Further we insist that the number of different values that appear in the matrix $A$ are from a finite set of values. For a family of matrices with describing circuits of size $g(n)$, we say the matrices are $g(n)$-succinct. (If not mentioned otherwise, we will call a matrix succinct if $g(n) = \text{poly}(n)$.)

Many matrices are succinct. For example Hadamard matrices are succinct, and so are products of two Hadamard matrices, for any $g(n)$ above some linear growth.

For now we subsume the distinction between input and ancilla qubits and suppose $m = n$, or at least $m = \Theta(n)$. We will overload notation and think of the rows and columns being indexed by $\{0,1\}^n$ or $[N] \stackrel{def}{=} \{0, \ldots, N-1\}$. Given a state vector $\mathbf{z} \in \mathbb{C}^N$ resulting from the computation and a set $S \subset [N]$ of coordinates corresponding to "accept" for a measurement in the standard basis, we wish to compute or approximate the quantity $\mu_S(\mathbf{z})$ from Definition 2.4, which gives the quantum acceptance probability.

**Definition 3.1.** *Let $S \subseteq [N]$ be succinct and have cardinality $|S| = N/r$ for some integer $r \geqslant 1$. Let $\mathbf{v} = (v_1, \ldots, v_r) \in \mathbb{C}^r$ be some fixed vector. An $N \times N$ matrix $M$ is called $(S, \mathbf{v})$-decomposable if the following property is satisfied. Let the row-submatrix of $M$ indexed by the rows in $S$ be given by $(M_0, \ldots, M_{r-1})$, where each $M_i$ is an $N/r \times N/r$ matrix. Then there exists a unitary $N/r \times N/r$ matrix $U$ such that for each $i \in [r]$, $M_i = v_i \cdot U$.*

We observe that provided $S$ is succinct, this definition allows us general range of $S$ when building a $D^3$ circuit. Namely, suppose the components $\mathcal{A}$ and $\mathcal{D}$ are originally designed with regard to a different measurement set $S'$ of the same cardinality. We can then surround them by permutation matrices carrying $S$ onto $S'$, and the result is still a $D^3$ circuit by Lemma 2.2. Thus it does not matter that our key definition is worded to depend on the set $S$, and there is freedom to choose an $S$. Note that the decomposability definition depends only on the corresponding matrix rows. We note that the sets $S$ resulting from important examples are always succinct, and this is a reasonable stipulation for measuring a quantum circuit of size $\text{poly}(n)$ in general.

We show that the set of decomposable matrices is non-empty by showing that two well-known families of unitary matrices are decomposable. The first, $H^n$, contributes depth only 1 in a quantum circuit, but the second is reckoned as log-depth, so our $D^3$ notion is incomparable with previous definitions of constant-depth quantum circuits.

**Hadamard Matrices.** We begin with the definition of Hadamard matrices:

**Definition 3.2.** *The Hadamard matrix $H^n$ of order $N = 2^n$ is defined recursively as follows*

$$H^n = \frac{1}{\sqrt{2}} \cdot \begin{pmatrix} H^{n-1} & H^{n-1} \\ H^{n-1} & -H^{n-1} \end{pmatrix},$$

where $H^0 = (1)$.

It is well-known that Hadamard matrices are unitary.

If one unravels the recursion above for $i$ steps, then every consecutive "chunk" of $N/2^i$ rows of $H^n$ are of the form

$$2^{-i/2} \left( a_0 \cdot H^{n-i} \ a_2 \cdot H^{n-i} \ldots a_{2^i-1} \cdot H^{n-i} \right),$$

where each $a_j \in \{-1, 1\}$. This implies that

**Proposition 3.3.** *Let $H^n$ be the $N \times N$ Hadamard matrix and let $S = \{j \cdot N/2^i, \ldots, (j+1)N/2^i - 1\}$ for some $i \geq 1$ and $j \in [2^i]$. Then $H^n$ is $(S, (v_0, \ldots, v_{2^i-1}))$-decomposable where for each $j \in [2^i]$, $v_j \in \{-2^{-i/2}, 2^{-i/2}\}$.*

**Fourier Matrices.** We begin with the definition of quantum Fourier matrices:

**Definition 3.4.** *The Fourier matrix $F^N$ of order $N$ has as its $(i,j)$th entry $\frac{1}{\sqrt{N}} \cdot \omega_N^{ij}$, where $\omega_N$ is the $N$th root of unity and $i, j \in [N]$.*

It is well-known that $F^N$ is a unitary matrix over $\mathbb{C}$.

Let $k$ divide $N$. Then it is easy to see that $\omega_N^k = \omega_{N/k}$ is the $N/k$th root of unity. Further, note that for any $i = k\ell$ for some $\ell \in [N/k]$, we have that the $i$th row of $F^N$ is given by

$$1/\sqrt{N} \cdot \left( 1, \ldots, \omega_N^{k\ell(N/k-1)}, \omega_N^{N/k \cdot k\ell}, \ldots, \omega_N^{k\ell(2N/k-1)}, \cdots, \omega_N^{(k-1)N/k \cdot k\ell}, \ldots, \omega_N^{(N-1)k\ell} \right),$$

which is the same as

$$1/\sqrt{N} \cdot \left( 1, \ldots, \omega_{N/k}^{\ell(N/k-1)}, 1, \ldots, \omega_{N/k}^{\ell(N/k-1)}, \cdots, 1, \ldots, \omega_{N/k}^{\ell(N/k-1)} \right).$$

In other words the $k\ell$th row in $F^N$ is the $\ell$th row of $F^{N/k}$ repeated $k$ times (and multiplied by a factor of $1/\sqrt{k}$). In other words, the rows of $F^N$ that are divisible by $k$ form the matrix

$$\frac{1}{\sqrt{k}} \left( \underbrace{F^{N/k} \ F^{N/k} \cdots F^{N/k}}_{k \text{ times}} \right).$$

This implies the following result:

**Proposition 3.5.** *Let $F^N$ be the Fourier matrix of order $N$. Let $k$ be an integer dividing $N$ and let $S = \{i \in [N] | i \mod (k) \equiv 0\}$. Then $F^N$ is $\left( S, \frac{1}{\sqrt{k}}(\underbrace{1, 1, \ldots, 1}_{k \text{ times}}) \right)$-decomposable.*

## 3.1 Instances of $D^3$ quantum algorithms

We will show soon why such matrices are interesting from the perspective of simulating a $D^3$ polynomial time quantum algorithm. For now we note:

1. In the Deutsch-Jozsa algorithm, $\mathcal{A}$ and $\mathcal{B}$ are $H^{n+1}$ and $H^n$ (with one ancilla qubit), while $\mathcal{D}$ is the deterministic function $f$ plus $y \oplus f(x)$ on the ancilla.

2. In Shor's algorithm, as presented in the first diagram on $3n$ qubits in [Bea03], $\mathcal{A}$ is $H^{2n}$ with constant-1 on the last $n$ qubits, $\mathcal{B}$ is the inverse Quantum Fourier Transform, and $\mathcal{D}$ are applications of $x \mapsto a^{2^j} x \bmod M$ on the last $n$ qubits, controlled from the first $2n$.

## 4 An Exact Characterization

In this section, we show that if each entry of a vector $\mathbf{x}$ takes some fixed discrete values, then there is an exact characterization of $\mu_S(M\mathbf{x}^T)$, where $M$ is $(S, \cdot)$-decomposable.

**Lemma 4.1.** *Let $\Sigma \subset \mathbb{C}$ be a finite set. Let $r \geq 1$ be an integer such that $r$ divides $N$. Let $S \subseteq [N]$ with $|S| = N/r$ and let $\mathbf{v} \in \mathbb{C}^r$ be a fixed vector. Let $M$ be an $N \times N$ matrix that is $(S, \mathbf{v})$-decomposable. Then for every $\mathbf{x} \in \Sigma^N$, we have*

$$\mu_S(M\mathbf{x}^T) = \sum_{\mathbf{c} \in \Sigma^r} \langle \mathbf{c}, \mathbf{v} \rangle^2 \cdot \#_{\mathbf{c}}(\mathbf{x}).$$

*Proof.* Let $U$ be the $N/r \times N/r$ unitary matrix such that the row sub-matrix of $M$ corresponding to $S$ is given by $(M_0, \ldots, M_{r-1})$ such that for each $i \in [r]$, $M_i = v_i \cdot U$. Further, let $\mathbf{x} = (\mathbf{x}^0, \ldots, \mathbf{x}^{r-1}) \in \left(\Sigma^{N/r}\right)^r$. It is easy to check that

$$(M\mathbf{x}^T)_S = M_0(\mathbf{x}^0)^T + \cdots + M_{r-1}(\mathbf{x}^{r-1})^T, \tag{1}$$

where for a vector $\mathbf{y} \in \mathbb{C}^N$, $\mathbf{y}_S$ denotes the vector projected to indices in $S$. Consider the following sequence of relationships:

$$\mu_S(M\mathbf{x}^T) = \|(M\mathbf{x}^T)_S\|_2^2 \tag{2}$$

$$= \|\sum_{i=0}^{r-1} M_i (\mathbf{x}^i)^T\|_2^2 \tag{3}$$

$$= \|U \cdot \left(\sum_{i=0}^{r-1} (v_i \cdot \mathbf{x}^i)\right)^T\|_2^2 \tag{4}$$

$$= \|\left(\sum_{i=0}^{r-1} (v_i \cdot \mathbf{x}^i)\right)^T\|_2^2 \tag{5}$$

$$= \sum_{j=0}^{N/r-1} \left(\sum_{i=0}^{r-1} v_i \cdot x_j^i\right)^2 \tag{6}$$

$$= \sum_{\mathbf{c} \in \Sigma^r} \langle \mathbf{c}, \mathbf{v} \rangle^2 \cdot \#_{\mathbf{c}}(\mathbf{x}). \tag{7}$$

In the above, (2) follow from definitions. (3) follows from (1). (4) follows from the definition of $M_i$'s. (5) follows from the fact that $U$ is a unitary matrix. (6) follows from stating $\mathbf{x}^j = (x_0^j, \ldots, x_{N/r-1}^j)$. Finally, (7) follows by re-arranging the sum, the fact that $\mathbf{x} \in \Sigma^N$ and by Definition 2.5. $\square$

## 5 Approximating $\mu_S(M\mathbf{x}^T)$

In this section, we show how to approximate $\mu_S(M\mathbf{x}^T)$ quickly using Lemma 4.1 when $\mathbf{x}$ takes values in a bounded set within the reals and is succinct.

Given the exact characterization from Lemma 4.1, the algorithm to approximate $\mu_S(M\mathbf{x}^T)$ for succinct $\mathbf{x}$ follows by the natural sampling algorithm. (The details are in Appendix A.)

**Theorem 5.1.** *Let $\varepsilon > 0$ be a real. Let $\Sigma \subset \mathbb{R}$ be a finite set and let $A = \max_{b \in \Sigma} |b|$. Let $r \geqslant 1$ be an integer such that $r$ divides $N$. Let $S \subseteq [N]$ with $|S| = N/r$ and let $\mathbf{v} \in \Gamma^r$ be a fixed vector, where $\Gamma \subset \mathbb{R}$ with $B = \max_{a \in \Gamma} |a|$. Let $M$ be an $N \times N$ matrix that is $(S, \mathbf{v})$-decomposable.*

*Then for every succinct $\mathbf{x} \in \Sigma^N$, there exists an $O\left(\frac{r^6}{\varepsilon^2} \cdot |\Sigma|^{2r} \log |\Sigma| \cdot (|A||B|)^4 \cdot \log^{O(1)} N\right)$-time randomized algorithm that outputs an estimate $\hat{\mu}$ such that with probability at least $2/3$,*

$$|\hat{\mu} - \mu_S(M\mathbf{x}^T)| \leqslant \varepsilon \cdot \frac{N}{r}.$$

Next, we state a corollary that will be useful.

**Corollary 5.2.** *Let $\varepsilon > 0$ be a real. Let $\Sigma \subset \mathbb{R}$ be a finite set and let $A = \max_{b \in \Sigma} |b|$. Let $r \geqslant 1$ be an integer such that $r$ divides $N$. Let $S \subseteq [N]$ with $|S| = N/r$ and let $\mathbf{v} \in \Gamma^r$ be a fixed vector, where $\Gamma \subset \mathbb{R}$ with $B = \max_{a \in \Gamma} |a|$. Let $M$ be an $N \times N$ matrix that is $(S, \mathbf{v})$-decomposable.*

*Let $|\Sigma|, |A|, |B| \leqslant O(1)$ and $r \leqslant O(\log \log N)$. Then for every succinct $\mathbf{x} \in \Sigma^N$, one can approximate $\mu_S(M\mathbf{x}^T)$ to within an additive $\varepsilon N/r$ factor in randomized $\mathrm{poly}(\log N)$ time.*

In particular, by setting $r = O(\log \log N) = O(\log n)$, Corollary 5.2 implies Theorem 1.1.

### 5.1 Sub-linear Time Fourier Transforms

We now compare our techniques with the existing ones on approximating Fourier coefficients. If $M$ is the Hadamard matrix (and the Fourier matrix resp.), then given $M \cdot \mathbf{x}^T$ is the Fourier transform of $\mathbf{x}$ over $\{0,1\}^N$ (and $\mathbb{C}^N$ resp.). Further, there exist sub-linear algorithms that can estimate any Fourier coefficient of $\mathbf{x}$ to within an additive $\gamma > 0$ error in time $\mathrm{poly}(\log N, \frac{1}{\gamma})$.[1] This was implicitly for done for the Hadamard matrix in the work of Goldreich and Levin [GL89]. The result for Fourier matrix was done by Gilbert et al. [GGI+02]. This was improved upon by Akavia et al. [AGS03]. (There are now deterministic algorithms known to estimate the Fourier coefficients [Aka10, Iwe08].)

To estimate $\mu_S(M\mathbf{x}^T)$ one could try and use the sub-linear algorithms above to estimate the Fourier coefficients indexed by $S$. There are two issues with this approach: (i) If one added the estimates of $\{(M\mathbf{x}^T)_i\}_{i \in S}$, then to get an overall error of $\varepsilon$ one would have to set $\gamma = O(\varepsilon/|S|)$– this would lead to a run time that is $\mathrm{poly}(|S|)$. (ii) Further, even if one had perfect estimates of the Fourier

---
[1]Actually, these algorithm can compute Fourier coefficients with mass $\tau \cdot \|M\mathbf{x}^T\|_2$ to within additive error $\varepsilon > 0$ for any $\tau, \varepsilon > 0$ in time $\mathrm{poly}(\log N, \frac{1}{\tau}, \frac{1}{\varepsilon})$. The claimed result follows by just setting $\tau = \varepsilon = \gamma$.

coefficients in $\{(M\mathbf{x}^T)_i\}_{i \in S}$, even adding them up would take $\Omega(|S|)$ time. In our simulations, we have $|S| \geqslant \tilde{\Omega}(N)$, which of course is too inefficient. Alternatively, one could use the fact that these estimation algorithms can estimate the Fourier coefficient with at least $\tau$ fraction of the total mass. However note that we can have $\mu_S(M\mathbf{x}^T)$ be $\Omega(1)$ while all the Fourier coefficients themselves have $O(1/|S|)$ of the mass. Since the sub-linear algorithms to estimate the top Fourier coefficient have polynomial dependence in $1/\tau$, this method can be prohibitively expensive.

It is perhaps instructive to see how our simulation avoids the pitfalls above. The most obvious reason is that Lemma 4.1 gives a direct characterization of the quantity $\mu_S(M\mathbf{x}^T)$ that we want to estimate. However, we do not compute $\mu_S(M\mathbf{x}^T)$ exactly but estimate is by a simple sampling procedure. In particular, we show that we estimate each of the summand in the equality in Lemma 4.1 with accuracy $\gamma > 0$ and then take the union bound. However, unlike the case in the previous paragraph, we have to take union bound over $\exp(N/|S|)$ events. In our simulations, we use $|S| \geqslant \Omega(N/\log\log N)$ which implies that we need to set $\gamma = \varepsilon \exp(-\frac{N}{|S|})$. Since our run time is poly$(1/\gamma)$, by our choice of $|S|$ we obtain an efficient poly$(\log N)$ runtime overall.

## 6 Open Problems

Our work raises several interesting questions. Can Theorem 1.1 be improved to allow the measurement of more qubits? We can increase the number of qubits to $O(\log n)$ giving only a $(1/3, 2/3)$ separation and still have the simulation take polynomial randomized time.

**Problem 1.** *What is the exact classical cost of simulating any $D^3$-algorithm that measures $r$ bits?*

Our current result shows that the cost is upper-bounded by $2^{O(r)}$ times the cost of running the circuit. We will investigate whether this bound can be improved.

We note that our techniques and those in the sub-linear algorithms to estimate Fourier coefficients are complementary: our results work better when $|S|$ is large where as other techniques work better when $|S|$ is small. An intriguing possibility is:

**Problem 2.** *Can we design a simulation algorithm for $D^3$ quantum polynomial time algorithms that has the best of both the worlds?*

As a technical matter, we may also ask where and why the $D^3$ definition may not allow standard transformations that establish robustness properties of quantum circuits, such as measuring just one qubit and/or deferring measurements. The strengthened succinctness notion is a non-uniform version of the idea of polynomial-time *rankability* [GS91, HR90] with different scaling, which may provide further scope for studying it.

The final open question is the possibility of a new hard function for cryptographic applications coming out of this work.

**Problem 3.** *Can we utilize the following as a hard problem for cryptographic applications: Given $r(n)$ (say $O(\log n)$) qubits from the quantum part of Shor's algorithm, compute the complete factorization.*

## A  Proof of Theorem 5.1

*Proof.* Let $\ell \geqslant 1$ be a parameter that will be fixed later to be $\ell = O(1/\varepsilon^2 \cdot |\Sigma|^{2r} \cdot r^5 \cdot (|A| \cdot |B|)^4 \cdot \log |\Sigma|)$. Consider the following simple sampling algorithm:

1. Set $p \leftarrow 0$.

2. Repeat $\ell$ times:
   - Pick a random $i \in [N/r]$.
   - Set $p \leftarrow p + \left(\sum_{j=0}^{r-1} v_j \cdot x_j^i\right)^2$.

3. Output $\delta \leftarrow \frac{p}{\ell}$.

We will argue shortly that $\delta$ approximates $\frac{\mu_S(M\mathbf{x}^T)}{N/r}$ to within an additive factor of $\varepsilon$. First, we bound the run time of the algorithm. It is easy to see that step 2 dominates the run time.[2] Each of the $\ell$ repetitions involves reading $r$ $x_j^i$ values (each of which takes poly$(\log N)$ time as $\mathbf{x}$ is succinct) and performing $O(r)$ additions and multiplications. Altogether, the algorithm takes $O(\ell(r + r \cdot \log^{O(1)} N))$ time, which with our choice of $\ell$ implies the claimed running time.

We now argue that the probability that $|\delta - \mu_S(M\mathbf{x}^T)/(N/r)| > \varepsilon$ is at most $1/3$. Assigning the estimate $\hat{\mu} = \delta \cdot N/r$ would then complete the proof.

---
[2]We'll make the simplifying assumption that all basic arithmetic operations are $O(1)$ time.

For every $\mathbf{c} \in \Sigma^r$, let $\delta_\mathbf{c}$ be the random variable denoting the fraction of the indices $i \in [N/r]$ chosen in Step 2 such that $(x_0^i, \ldots, x_{r-1}^i) = \mathbf{c}$. Note that

$$\delta = \sum_{\mathbf{c} \in \Sigma^r} \delta_\mathbf{c} \langle \mathbf{c}, \mathbf{v} \rangle^2. \tag{8}$$

We will show that for every $\mathbf{c} \in \Sigma^r$ with probability at most $\frac{1}{3 \cdot |\Sigma|^r}$,

$$|\delta_\mathbf{c} - \frac{\#_\mathbf{c}(\mathbf{x})}{N/r}| > \frac{\varepsilon}{|\Sigma|^r \cdot (r|A||B|)^2}. \tag{9}$$

Since $\langle \mathbf{c}, \mathbf{v} \rangle \leqslant r|A||B|$, we have with probability at most $\frac{1}{3 \cdot |\Sigma|^r}$,

$$|\langle \mathbf{c}, \mathbf{v} \rangle^2 \cdot \delta_\mathbf{c} - \frac{\#_\mathbf{c}(\mathbf{x}) \cdot \langle \mathbf{c}, \mathbf{v} \rangle^2}{N/r}| > \frac{\varepsilon}{|\Sigma|^r}. \tag{10}$$

Taking union bound over all the choice of $\mathbf{c}$, (10), (8) and Lemma 4.1 implies that with probability at most $1/3$,

$$|\delta - \frac{\mu_S(M\mathbf{x}^T)}{N/r}| > \varepsilon.$$

as desired.

To complete the proof, we argue (9). First, note that

$$\mathbf{E}[\delta_\mathbf{c} \ell] = \frac{\#_\mathbf{c}(\mathbf{x}) \ell}{N/r}.$$

Thus, by the (additive) Chernoff bound, the probability that $|\delta_\mathbf{c} \ell - \mathbf{E}[\delta_\mathbf{c} \ell]| > \alpha \ell$ is upper bounded by

$$\exp(-2\alpha^2 \ell) \leqslant \frac{1}{3 \cdot |\Sigma|^r},$$

where the inequality follows if

$$\ell \geqslant \frac{1}{2\alpha^2} \cdot \log(3|\Sigma|^r).$$

Noting that to prove (9), we need to pick $\alpha = \frac{\varepsilon}{|\Sigma|^r \cdot (r|A||B|)^2}$, we need

$$\ell \geqslant \frac{|\Sigma|^{2r}}{2\varepsilon^2} \cdot (r|A||B|)^4 \log(3|\Sigma|^r),$$

which is satisfied if we pick

$$\ell = O\left(1/\varepsilon^2 \cdot |\Sigma|^{2r} \cdot r^5 \cdot (|A| \cdot |B|)^4 \cdot \log |\Sigma|\right).$$

$\square$